\documentclass[
prd,
twocolumn,
nofootinbib,
floatfix,
showpacs ,amssymb, aps]{revtex4}
\usepackage{amsmath}
\usepackage{color}

\usepackage{comment}
\usepackage{graphicx}

\usepackage{amsmath,amssymb}
\usepackage{bm}
\usepackage{enumitem}

\usepackage{natbib}
\usepackage{wasysym}

\newcommand{\F}{\mathcal{F}}

\begin{document}

\title{Stochastic gravitational wave background from hydrodynamic turbulence in differentially rotating neutron stars}

\author{Paul D. Lasky}
	\email{paul.lasky@unimelb.edu.au}
\author{Mark F. Bennett}
\author{Andrew Melatos}
\affiliation{School of Physics, University of Melbourne, Parkville, VIC 3010, Australia}

		\begin{abstract}
			Hydrodynamic turbulence driven by crust-core differential rotation imposes a fundamental noise floor on gravitational wave observations of neutron stars.  The gravitational wave emission peaks at the Kolmogorov decoherence frequency which, for reasonable values of the crust-core shear, $\Delta\Omega$, occurs near the most sensitive part of the frequency band for ground-based, long-baseline interferometers.  We calculate the energy density spectrum of the stochastic gravitational wave background from a cosmological population of turbulent neutron stars generalising previous calculations for individual sources.  The spectrum resembles a piecewise power law, $\Omega_{\rm gw}(\nu)=\Omega_{\alpha}\nu^{\alpha}$, with $\alpha=-1$ and $7$ above and below the decoherence frequency respectively, and its normalisation scales as $\Omega_{\alpha}\propto\left(\Delta\Omega\right)^{7}$.  Non-detection of a stochastic signal by Initial LIGO implies an upper limit on $\Delta\Omega$ and hence by implication on the internal relaxation time-scale for the crust and core to come into co-rotation, $\tau_{d}=\Delta\Omega/\dot{\Omega}$, where $\dot{\Omega}$ is the observed electromagnetic spin-down rate, with $\tau_{d}\lesssim 10^{7}\,{\rm yr}$ for accreting millisecond pulsars and $\tau_{d}\lesssim 10^{5}\,{\rm yr}$ for radio-loud pulsars.  Target limits on $\tau_{d}$ are also estimated for future detectors, namely Advanced LIGO and the Einstein Telescope, and are found to be astrophysically interesting.
		\end{abstract}
		
		\pacs{95.85.Sz 
		04.30.Db	
		97.60.Jd	
		}
		
		
		
		
		\maketitle

\section{Introduction}

The electromagnetic braking torque acting on a neutron star drives persistent differential rotation between the rigid crust and the multiple superfluid components in the interior, causing observed phenomena like rotational glitches \cite{lyne06,melatos08,espinoza11a}.  Classical relaxation processes, like Ekman pumping and Sweet-Parker circulation, and quantum mechanical relaxation processes, like superfluid vortex creep, determine the long-term angular velocity shear between the various internal components \cite{anderson75,easson79,warszawski11,melatos12}.  In turn, differential rotation drives turbulence when the Reynolds number is high ($\rm Re\gtrsim10^{11}$) as in a neutron star \cite{greenstein70,melatos10}.  The turbulence takes two distinct forms: (1) a Kolmogorov-like cascade of macroscopic ``eddies'' or circulation cells \cite{peralta05,peralta06a,peralta06b,peralta08,melatos07,melatos10}; and (2) a self-sustaining tangle of microscopic quantised vortices \cite{tsubota09,paoletti11}, created when the rectilinear vortex array in a uniformly rotating superfluid is disrupted by instabilities driven by meridional circulation \cite{peralta06b,melatos07}, interfacial and bulk two-stream instabilities \cite{andersson03a,mastrano05,andersson07} and nuclear pinning forces \cite{link12a,link12b}.

Turbulence driven by differential rotation is axisymmetric when averaged over long times but non-axisymmetric instantaneously.  It therefore emits stochastic gravitational radiation.  The root-mean-square gravitational wave strain assuming incompressibility and long-term isotropy is given by \cite{melatos10}
\begin{align}
	h_{\rm rms}=\,&5\times10^{-28}\left(\frac{M_{\star}}{1.4\,M_{\odot}}\right)\left(\frac{R_{\star}}{10\,{\rm km}}\right)^{3}\notag\\
		&\times\left(\frac{d}{1\,{\rm kpc}}\right)^{-1}\left(\frac{\Delta\Omega}{10\,{\rm rad\,s}^{-1}}\right)^{3},\label{hrms}
\end{align}   
where $M_{\star}$ and $R_{\star}$ are respectively the mass and radius of the neutron star, $d$ is the distance from Earth and $\Delta\Omega$ is the angular velocity difference between the slower crust (angular velocity $\Omega$) and faster core.  The emission is predicted to be astrophysically relevant [and potentially detectable by third-generation interferometers like the Einstein Telescope (ET)] for sources including protoneutron stars \cite{ott06}, accreting millisecond pulsars, accreting white dwarfs on the verge of accretion-induced collapse \cite{dessart07,metzger08} and young pulsars with super-rotating cores, whose deceleration is inhibited by buoyancy \cite{melatos12}.  Even before any gravitational-wave detections, the indirect spin-down limit from radio timing observations puts interesting upper limits on $\Delta\Omega$ (see the left panel of figure 5 in Ref. \cite{melatos10}), with $\Delta\Omega/\Omega\lesssim10^{-2}$ in some sources, approaching the shear inferred from glitch data. 

In this paper we extend the single-source calculations in Ref. \cite{melatos10} to calculate the stochastic gravitational wave signal coming from all neutron stars in the Universe.  We then use existing non-detections by the Laser Interferometer Gravitational-wave Observatory (LIGO) to place an upper limit on $\Delta\Omega$ across the cosmological population as a whole, generalising the single-source limits.  Many kinds of sources contribute to the stochastic background in the LIGO frequency band; for reviews of various sources see Refs. \cite{maggiore00,regimbau08,regimbau11}.  Searches from the first generation of LIGO-class detectors have already pushed the gravitational wave energy density, $\Omega_{\rm gw}(\nu)$, at frequency $\nu$, below that inferred from Big Bang nucleosynthesis (BBN) and the cosmic microwave background \cite{abbott09}.  Cross-correlation searches are performed assuming a power law of the form $\Omega_{\rm gw}(\nu)=\Omega_{\alpha}\nu^{\alpha}$; they place an upper bound on $\Omega_{\alpha}$ given $\alpha$  \cite{abbott05,abbott07a,abbott07c,abadie12}.  We show in this paper that the gravitational-wave energy density from turbulent neutron stars is well approximated by a piecewise power law with a rising component ($\alpha=7$) at low $\nu$ and a decaying tail ($\alpha=-1$) at high $\nu$.  The power laws join and peak near the turbulence decoherence frequency, with $\Omega_{\rm gw}(\nu)$ at the peak scaling with the seventh power of $\Delta\Omega$.

The structure of the paper is as follows.  In section \ref{sec2} we review the relevant statistical properties of superfluid turbulence and calculate the gravitational wave signal from a single turbulent neutron star (\ref{singlesource}) and multiple stars (\ref{multiplesource}).  In section \ref{upperlimits} we use LIGO non-detections to derive upper limits on $\Delta\Omega$ under three scenarios: a universal $\Delta\Omega$ set by nuclear physics (\ref{unique}), a broad $\Delta\Omega$ distribution set by the balance between internal damping and electromagnetic spin down (\ref{radioloud}), and a narrow $\Delta\Omega$ distribution for the population of accreting millisecond pulsars (\ref{msp}).  The prospects for direct detection of this cosmological background by current and future detectors are summarised in section \ref{conclusion}.

\section{Gravitational Radiation from Neutron Star Turbulence}\label{sec2}

\subsection{Turbulence statistics}\label{SSINS}
We first review briefly the main statistical properties (e.g. the autocorrelation function) of the stochastic gravitational wave signal emitted by superfluid turbulence in a single, differentially rotating neutron star.  We refer the reader to Ref. \cite{melatos10} for details of the derivation.  The stochastic background resulting from the superposition of multiple sources at multiple redshifts is calculated in section \ref{multiplesource}.

The wave strain in the transverse-traceless gauge as measured by an observer at a distance $d$ is given by
\begin{align}
	h_{jk}^{\rm TT}=\frac{G}{c^{5}d}\sum_{\ell=2}^{\infty}\sum_{m=-\ell}^{\ell}\frac{\partial^{\ell}S^{\ell m}(t)}{\partial t^{\ell}}T^{B2,\ell m}_{jk},\label{hTT}
\end{align}
where $S^{\ell m}$ is the $\left(\ell,m\right)$ current multipole written as a function of the retarded time, $t$, and $T^{B2,\ell m}_{jk}$ is the tensor spherical harmonic describing the angular dependence of the radiation field (see equation 2.30f of Ref. \cite{thorne80}).  In the Newtonian approximation (i.e. slow internal motions and weak internal gravity), which is adequate for describing subsonic turbulence driven by slow differential rotation (i.e. $\Delta\Omega/\Omega\ll1$), the current multipole is
\begin{align}
	S^{\ell m}=\,&-\frac{32\pi}{\left(2\ell+1\right)!!}\left[\frac{\ell+2}{2\ell\left(\ell-1\right)\left(\ell+1\right)}\right]^{1/2}\notag\\
		&\times\int d^{3}{\bf x}\,r^{\ell}{\bf x}\cdot{\rm curl}\left(\rho{\bf v}\right)Y^{\ell m\star},\label{Slm}
\end{align}
where $Y^{\ell m}$ is a scalar spherical harmonic, ${\bf v}({\bf x},t)$ is the turbulent velocity field and $\rho({\bf x})$ is the fluid mass density.  For incompressible turbulence, the mass multipoles vanish to a good approximation, as the turbulent motions  are subsonic and $\rho({\bf x})$ is uniform\footnote{Density perturbations are of order $\rho R_{\star}^{2}\left(\Delta\Omega\right)^{2}/c_{s}^{2}$, where $c_{s}$ is the sound speed.  Hence the mass quadrupole is smaller by a factor $R_{\star}\Delta\Omega c/c_{s}^{2}$ than the current quadrupole.  See \cite{zrake12} for recent high-resolution simulations of compressible, relativistic turbulence.}.  The gravitational wave signal is dominated by $\ell=2$ for most realistic neutron stars, i.e. objects with Reynolds number ${\rm Re}\lesssim\left(c/R_{\star}\Delta\Omega\right)^{8}$ \cite{melatos10}.


We consider stationary, isotropic turbulence, for which the mean wave strain at the observer vanishes, and the leading non-zero moment is the autocorrelation function,
\begin{align}
	C(\tau)=\left<h_{jk}^{\rm TT}(t)h_{jk}^{\rm TT}(t')^{\star}\right>,\label{Ctau}
\end{align}
with $\tau=t'-t$, where $\left<\ldots\right>$ denotes the ensemble average over realisations of the turbulence.  From equation (\ref{Slm}), the mean-square wave strain, $h_{\rm rms}^{2}=C(0)$, is proportional to the zero-lag autocorrelation of the fluctuating vorticity field, ${\rm curl}\,{\bf v}({\bf x}, t)$.

Three-dimensional, global simulations of shear-driven neutron star turbulence suggest that, for a two-component Hall--Vinen--Bekarevich--Khalatnikov superfluid, the flow is approximately isotropic and stationary for ${\rm Re}\gtrsim10^{4}$ \cite{peralta05,peralta06b,melatos07,peralta08}.  We adopt the standard Kraichnan-Kolmogorov form for the unequal-time velocity correlator \cite{kraichnan59,kosowsky02,melatos10}\footnote{Equation (\ref{vv}) differs from equation (2) of Ref. \cite{melatos10} by a factor of $(2\pi)^{3}$, which we have absorbed into the definition of the power spectral density, $P(k)$.}
\begin{align}
	\left<v_{m}\left({\bf k},t\right)v_{q}\left({\bf k}',t'\right)^{\star}\right>=V\hat{P}_{mq}(k)F\left(k,\tau\right)\delta\left({\bf k}-{\bf k}'\right).\label{vv}
\end{align}
Here, $V$ is the total volume of the system, $\hat{P}_{mq}(k)=\delta_{mq}-\hat{k}_{m}\hat{k}_{q}$ is a projection operator perpendicular to the wave vector ${\bf k}$, $P(k)$ is the power spectral density [i.e. $(2\pi)^{-3}k^{2}P(k)$ is the kinetic energy in the flow per unit wavenumber], and we have
\begin{align}
	F(k,\tau)=P(k){\rm exp}\left[-\pi\eta(k)^{2}\tau^{2}/4\right],
\end{align}
with 
\begin{align}
	\eta(k)=\left(2\pi\right)^{-1/2}\epsilon^{1/3}k^{2/3},
\end{align}
where $\eta(k)^{-1}$ is the eddy turnover time at wavenumber $k$, and $\epsilon$ is the energy dissipation rate per unit enthalpy.

The power spectral density is not known for superfluid turbulence.  In the absence of specific information, we assume the Kolmogorov law, $P(k)\propto k^{\alpha}$, with $\alpha=-11/3$, which characterises isotropic, high-Re, Navier-Stokes turbulence.  The power law extends between the stirring wavenumber, $k_{s}=2\pi/R_{\star}$, where $R_{\star}$ is the stellar radius, and the viscous dissipation wavenumber, $k_{d}=\left[8\epsilon/\left(27v^{3}\right)\right]^{1/4}$, where $v$ is the kinematic viscosity.  The Kolmogorov law is consistent with preliminary numerical simulations in the context of the two-component Hall--Vinen--Bekarevich--Khalatnikov model with $10^{4}\lesssim{\rm Re}\lesssim10^{5}$ in the viscous component \cite{melatos10}.  However, $P(k)$ remains unknown in both stratified Navier-Stokes turbulence and unstratified superfluid turbulence even in terrestrial experiments \cite{salort10}.  The subtle spatio-temporal anisotropies caused by layering, intermittency and rotational polarisation (e.g. of the superfluid vortex tangle) are discussed briefly in Appendix \ref{PowerSpecApp}, with pointers to the voluminous literature discussing these issues.

Combining the above equations, and expanding the plane-wave Fourier components (\ref{vv}) in spherical harmonics to evaluate $S^{2m}$, one arrives at the following formula for the autocorrelation function for the mode $(\ell,m)=(2,m)$:
\begin{align}
	\frac{C(\tau)}{h_{\rm rms}^{2}}=&\left[1-\frac{7\pi\eta(k_{s})^{2}\tau^{2}}{2}\right]{\rm exp}\left[-\frac{\pi\eta(k_{s})^{2}\tau^{2}}{4}\right]\notag\\
		&+2\pi^{2}\eta(k_{s})^{3}\tau^{3}\Big\{{\rm Erf}\left[\frac{\pi^{1/2}\eta(k_{s})\tau}{2}\left(\frac{k_{d}}{k_{s}}\right)^{2/3}\right]\notag\\
		&-{\rm Erf}\left[\frac{\pi^{1/2}\eta(k_{s})\tau}{2}\right]\Big\}.\label{autoC}
\end{align}
In Ref. \cite{melatos10} only the $\ell=m=2$ mode was considered. In reality all the modes $\ell=2$, $|m| \leq 2$ contribute to the wave strain. For a single source, the angular dependence of the combined wave strain is complicated. However one can show that all the modes $\ell=2$, $|m| \leq 2$ contribute equally to the total autocorrelation function, which is the quantity of interest for a stochastic background from many isotropically distributed sources; see equation (\ref{spherical}) below.

The decoherence time corresponding to the half-strain point, $C(\tau_{c})=h_{\rm rms}^{2}/2$, is
\begin{align}
	\tau_{c}&=0.35\eta(k_{s})^{-1}\\
		&=26\left(\frac{\Delta\Omega}{10\,{\rm rad}\,{\rm s}^{-1}}\right)^{-1}\,{\rm ms},\label{dec}
\end{align}
where $\Delta\Omega=\epsilon^{1/3}R_{\star}^{-2/3}$ is the angular velocity lag between crust and core.  

Equation (\ref{autoC}) is derived assuming $|T^{B2,2m}_{jk}|^{2}=1$, which is true only for an optimally situated observer; that is, $C(\tau)$ in equation (\ref{autoC}) is the maximum of equation (\ref{Ctau}) over observer orientation for a given $m$.  As we are ultimately interested in the stochastic background from an isotropic population of cosmological sources (section \ref{multiplesource}), it is more appropriate to replace $|T^{B2,2m}_{jk}|^{2}=1$ with the sky-averaged tensor harmonic product
\begin{align}
	\frac{1}{4\pi}&\int_{-1}^{1}d\left(\cos\theta\right)\int_{0}^{2\pi}d\phi\,\left<\sum_{m=-2}^{2}T_{jk}^{B2,2 m}\left.T_{jk}^{B2,2m}\right.^{\star}\right>=\frac{5}{4\pi},\label{spherical}
\end{align}
summed over the repeated indices $j,k$, where $\phi$ and $\theta$ represent latitude and longitude of the Earth relative to the source.  Hence equation (\ref{autoC}) must be multiplied by $5/(4\pi)$ in what follows.

\subsection{Single sources}\label{singlesource}
A key ingredient in calculating the stochastic background from multiple sources is the energy spectrum emitted by a single source.  To calculate the radiated energy per unit area per unit frequency in terms of $C(\tau)$, we begin with the standard form of the radiated energy per unit area per unit time expressed in the transverse-traceless gauge \cite[e.g.][]{misner73}
\begin{align}
	\frac{d^{2}E_{\rm gw}}{dSdt}=\frac{c^{3}}{16\pi G}\left<\frac{\partial h_{jk}^{\rm TT}(t)}{\partial t}\frac{\partial h_{jk}^{\rm TT}(t)}{\partial t}\right>.\label{lum}
\end{align}
The ensemble average in equation (\ref{lum}) is over realisations of the turbulence, as in equation (\ref{Ctau}).  Integrating both sides of equation (\ref{lum}) with respect to retarded time, we apply Parseval's theorem to obtain \cite{warszawski12}
\begin{align}
	\frac{d^{2}E_{\rm gw}}{dSd\nu_{e}}=\frac{c^{3}}{16\pi^{2}G}\left<\F\left[\frac{\partial h_{jk}^{\rm TT}}{\partial t}\right]\F\left[\frac{\partial h_{jk}^{\rm TT}}{\partial t}\right]\right>,\label{parc}
\end{align}
where $\F\left[\ldots\right]$ denotes the Fourier transform.  The above quantity is evaluated in the emitted (i.e. comoving) frame, remembering that the sources are at redshift $z\geq0$.  To clearly distinguish the comoving and observers frames, we denote the frequency emitted (observed) with (without) a subscript '$e$', such that $\nu_{e}=\nu(1+z)$.

The Wiener-Khintchine theorem relates the spectral density to the inverse Fourier transform of the autocorrelation function (eg., \cite{pottier10}).  Employing the identity \cite{gogoberidze07}
\begin{align}
	\left<\frac{\partial X(t)}{\partial t}\frac{\partial X(t')}{\partial t'}\right>=-\frac{d^{2}}{d\tau^{2}}\left<X(t)X(t')\right>,\label{ensemble}
\end{align}
valid for any {\it stationary}, differentiable random variable, $X(t)$, we obtain
\begin{align}
	\frac{d^{2}E_{\rm gw}}{dSd\nu_{e}}=-\frac{c^{3}T}{16\pi^{2}G}\int_{-\infty}^{\infty}d\tau\,\frac{d^{2}C(\tau)}{d\tau^{2}}{\rm e}^{i\nu_{e}\tau},\label{final}
\end{align}
where $T$ is the emitting lifetime.  The emitted energy increases monotonically with time while the source is on.  In equation (\ref{final}), the Wiener-Khintchine theorem is evaluated in the limit where $T$ is finite but much greater than $\tau_{c}$; the details are presented in Appendix \ref{app}.  

\begin{figure}
	\begin{center}
	\includegraphics[angle=0,width=0.95\columnwidth]{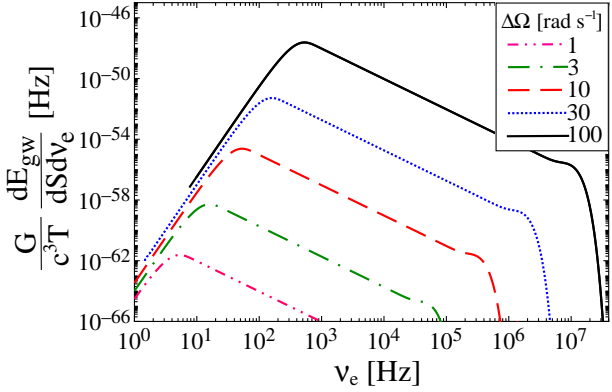}
	\end{center}
	\caption{\label{dEdSdnu} Gravitational wave energy per unit area per unit frequency emitted by a single neutron star as a function of frequency, for various values of the angular velocity shear $\Delta\Omega$.  
	}
\end{figure}

Twice differentiating the autocorrelation function given by equation (\ref{autoC}) and taking the Fourier transform according to equation (\ref{final}), we reach the final result
\begin{widetext}
\begin{align}
	\frac{d^{2}E_{\rm gw}}{dSd\nu_{e}}=\frac{c^{3}Th_{\rm rms}^{2}}{16\pi^{2}G}&\Bigg\{\left[
		\frac{28\nu_{e}^{2}}{\eta_{s}}\left(\frac{\nu_{e}^{2}}{\pi\eta_{s}^{2}}-\frac{3}{7}\right)
		-\frac{32}{\pi\eta_{s}^{3}\nu_{e}^{2}}\left(
			\frac{3}{4}\pi^{3}\eta_{s}^{6}+\frac{3}{4}\pi^{2}\eta_{s}^{4}\nu_{e}^{2}+\nu_{e}^{6}\right)\right]{\rm exp}\left(\frac{-\nu_{e}^{2}}{\pi\eta_{s}^{2}}\right)\notag\\
		&+\frac{32}{\pi\eta_{s}^{3}\nu_{e}^{2}}\left(\frac{k_{s}}{k_{d}}\right)^{4}\left[\frac{3}{4}\pi^{3}\eta_{s}^{6}\left(\frac{k_{d}}{k_{s}}\right)^{4/3}+\frac{3}{4}\pi^{2}\eta_{s}^{4}\nu_{e}^{2}\left(\frac{k_{d}}{k_{s}}\right)^{8/3}+\nu_{e}^{6}\right]
	{\rm exp}\left[\frac{-\nu_{e}^{2}}{\pi\eta_{s}^{2}}\left(\frac{k_{s}}{k_{d}}\right)^{4/3}\right]\Bigg\}
	\label{analytic}
\end{align}
\end{widetext}
with $\eta_{s}=\eta(k_{s})$.  This function is plotted in normalised form in figure \ref{dEdSdnu} for a typical neutron star with $M_{\star}=1.4\,M_{\odot}$, $R_{\star}=10\,{\rm km}$, kinematic viscosity, $v=1\,{\rm m}^{2}{\rm\,s}^{-1}$ and for various $\Delta\Omega$ values\footnote{We scale $v$ to $1\,{\rm m}^{2}{\rm s}^{-1}$, instead of the standard $10\,{\rm m}^{2}{\rm s}^{-1}$ \cite{cutler87} because Landau damping by weakly screened transverse plasmons \cite{shternin08} lowers $v$ by nearly an order of magnitude relative to the standard electron--electron scattering value.  The viscosity only affects the highest and lowest frequencies, and therefore plays an insignificant role in determining the integrated stochastic background.}.  The spectrum peaks near the inverse of the decoherence time, $\tau_{c}$, given by equation (\ref{dec}).  From equations (\ref{hrms}), (\ref{dec}) and (\ref{analytic}), one can show that the peak energy per unit area per unit frequency scales as $\Delta\Omega^{7}$, with $d^{2}E_{\rm gw}/dSd\nu_{e}\propto\nu_{e}^{6}$ for $\nu_{e}<1/\tau_{c}$ and $\propto\nu_{e}^{-2}$ for $\nu_{e}>1/\tau_{c}$.

\subsection{Multiple sources}\label{multiplesource}
To calculate the total energy emitted by multiple, continuously emitting sources at multiple redshifts $z_{e}$, consider an infinitesimal time interval $dt(z_{e})=(dt/dz_{e})dz_{e}$ between lookback times $t(z+dz_{e})$ and $t(z_{e})$.  The energy per unit area per unit logarithmic frequency interval emitted during this time interval equals 
\begin{align}
	\frac{\nu_{e}}{1+z_{e}}\frac{d^{2}E_{\rm gw}}{dSd\nu_{e}},
\end{align}
where $d^{2}E_{\rm gw}/dSd\nu_{e}$ is given by equation (\ref{final}), with $T$ replaced by $dt(z_{e})$,  the factor $(1+z_{e})^{-1}$ accounts for the redshifted energy of the gravitons, and $\nu_{e}=\nu(1+z_{e})$.  The number density of sources emitting during the time interval equals the total number of neutron stars born at redshifts $z_{b}\ge z_{e}$,
\begin{align}
	N(z_{e})=\frac{1}{M_{\star}}\int_{z_{e}}^{z_{\rm max}}dz_{b}\frac{dt}{dz_{b}}\frac{\dot{\rho}_{\star}(z_{b})}{1+z_{b}}\int_{M_{\rm min}}^{M_{\rm max}}dM\Phi(M),\label{numberdensity}
\end{align}
where $\dot{\rho}_{\star}(z_{b})$ is the star formation rate per unit comoving volume, 
$\Phi(M)$ is the initial mass function, $(M_{\rm min},\,M_{\rm max})$ defines the initial mass function range that forms neutron stars, $M_\star=1.4\,M_{\odot}$ is the neutron star mass, and $dt/dz_{b}$ is set by the cosmology.  Throughout the article we adopt a concordance cosmology with $\Omega_{m}=0.26$, $\Omega_{\Lambda}=0.74$ and $H=73\,{\rm km\,s^{-1}\,Mpc^{-1}}$.  We allow for the possibility that $d^{2}E_{\rm gw}/dSd\nu_{e}$ is also a function of $z_{b}$ through $\Delta\Omega$, which decreases as the source spins down and is therefore a function of its age.  Following convention, we can then express the energy density in the gravitational wave stochastic background, $\rho_{\rm gw}c^{2}$, as a fraction of the closure energy density per logarithmic frequency interval,
\begin{align}
	\Omega_{\rm gw}(\nu)&=\frac{1}{\rho_{c}c^{2}}\frac{d\left(\rho_{\rm gw}c^{2}\right)}{d\ln\nu}\label{OmegaA}\\
		&=\frac{1}{\rho_{c}c^{2}}\int_{0}^{\infty}dz_{e}\frac{N(z_{e})}{1+z_{e}}\left.\left(\nu_{e}\frac{dE_{\rm gw}}{d\nu_{e}}\right)\right|_{\nu_{e}=\nu\left(1+z_{e}\right)}.\label{Omega}
\end{align}
In (\ref{Omega}), $\rho_{c}c^{2}=3H_{0}^{2}c^{2}/8\pi G$ is the critical energy density required to close the Universe and $\nu$ is the frequency in the observer's frame.  

Superficially, equation (\ref{numberdensity}) and (\ref{Omega}) look identical to equation (5) of Ref. \cite{phinney01}.  Physically, however, the factors in the integrand have different interpretations, because our sources emit continuously, while those in Ref. \cite{phinney01} are discrete, short-lived bursts (i.e. inspirals).  In Ref. \cite{phinney01}, $N(z_{e})dz_{e}$ equals the infinitesimal number of inspiral events occurring between $z_{e}$ and $z_{e}+dz_{e}$, while $dE_{\rm gw}/d(\ln\nu_{e})$ is the total, non-infinitesimal energy per logarithmic frequency emitted by each event.  In contrast, in (\ref{Omega}), $N(z_{e})$ is the non-infinitesimal number of continuously emitting neutron stars in existence between $z_{e}$ and $z_{e}+dz_{e}$, while $dE_{\rm gw}/d(\ln\nu_{e})$ equals the infinitesimal energy per logarithmic frequency emitted during the time interval $dt(z_{e})=(dt/dz_{e})dz_{e}$.

From equations (\ref{analytic}) and (\ref{Omega}), one can show that the stochastic background is described approximately by a piecewise power law, $\Omega_{\rm gw}(\nu)=\Omega_{\alpha}\nu^{\alpha}$, with $\alpha=7$ and $\alpha=-1$ for $\nu<\nu_{c}$ and $\nu>\nu_{c}$ respectively.  The peak frequency, $\nu_{c}$, is a population-weighted average of the reciprocal of the decoherence time given in (\ref{dec}) (see section \ref{upperlimits}).

For the remainder of the article we adopt the modified Salpeter A initial mass function and the corresponding parametric fit for the star formation rate from \citet{hopkins06}.  For safety, we also verify our calculations against the star formation rate given in \citet{cucciati12}; the results are similar, with $\Omega_{\rm gw}(\nu_{c})$ differing by $\le6\%$ between the two mass functions.  As a rule, we take the range of zero-age main sequence masses that form neutron stars to be $8\le M/\,M_{\odot}\le 40$.  Varying the minimum and maximum masses over the ranges $4\le M_{\rm min}/M_{\odot}\le8$ and $20\le M_{\rm max}/M_{\odot}\le40$ respectively changes $\Omega_{\rm gw}(\nu_{c})$ by $\le25\%$, one of the smaller uncertainties in our overall calculation.  The initial mass range for neutron star formation is discussed more fully in Ref. \cite{oconnor11}

\section{Maximum shear from gravitational wave non-detections}\label{upperlimits}
The peak gravitational wave energy density, $\Omega_{\rm gw}(\nu_{c})$, from individual neutron stars scales as the seventh power of $\Delta\Omega$ as shown in section \ref{singlesource}.  However, detailed first-principles predictions for $\Delta\Omega$ are not yet available, nor is there any compelling observational support (e.g. from radio pulsar timing) for any particular choice of $\Delta\Omega$.  For the moment, therefore, we are obliged to use gravitational wave non-detections to place an upper limit on $\Delta\Omega$ across the neutron star population.  We do this for three illustrative astrophysical scenarios in this section:
\begin{itemize}
	\item a unique $\Delta\Omega$ value across the population, as expected if the shear is set by the balance between Magnus and nuclear pinning forces in the inner crust superfluid;
	\item a broad $\Delta\Omega$ distribution, wherein the shear is approximately proportional to the observed spin-down rate, balanced by some sort of internal relaxation (e.g. vortex creep, viscous damping or Sweet-Parker circulation); and
	\item a narrow $\Delta\Omega$ distribution at $\Delta\Omega\gtrsim10\,{\rm rad\,s}^{-1}$, due to fast accretion-driven spin-up in accreting millisecond rotators like low-mass X-ray binaries.
\end{itemize}

\subsection{Unique shear}\label{unique}
\begin{figure}
	\begin{center}
	\includegraphics[angle=0,width=0.95\columnwidth]{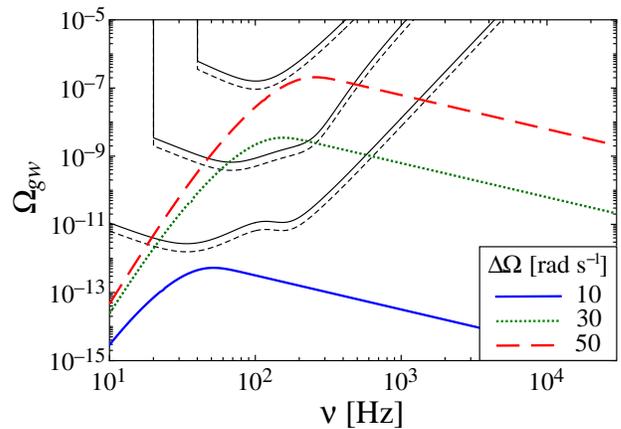}
	\end{center}
	\caption{\label{OmegaGWa}Gravitational wave energy density, $\Omega_{\rm gw}$, as a function of frequency, $\nu$, for three values of the angular velocity shear, $\Delta\Omega=10\,{\rm rad\,s}^{-1}$ (solid blue curve), $30\,{\rm rad\,s}^{-1}$ (dotted green curve) and $50\,{\rm rad\,s}^{-1}$ (dashed red curve).  The three solid (dashed) black curves are the one- (three-) year noise curves for Initial LIGO, Advanced LIGO and ET, running from top to bottom. }
\end{figure}

We begin by calculating $\Omega_{\rm gw}(\nu)$ assuming a constant shear, $\Delta\Omega$, in every neutron star in the Universe.  In figure \ref{OmegaGWa} we plot $\Omega_{\rm gw}$ as a function of $\nu$ for $\Delta\Omega=10$, $30$ and $50\,{\rm rad\,s}^{-1}$.  We also show the gravitational wave detection limits for Initial LIGO, Advanced LIGO and the proposed Einstein Telescope (ET), each for one and three years of data collection  (thick solid and dashed black curves respectively).  The sensitivity curves are taken from Ref. \cite{sathyaprakash09}, where an analytic fitting formula is given for the noise power spectral density of each detector (see table 1 of Ref. \cite{sathyaprakash09}).  The conversion to $\Omega_{\rm gw}$ for two co-located detectors and uncorrelated instrumental noise is given in section 8.1.2 of Ref. \cite{sathyaprakash09}.  Throughout this article we define a signal as being detectable if the predicted signal amplitude lies above the noise curve for that particular instrument at any frequency.  In reality, a more rigorous cross-correlation analysis will need to be done to detect a source.

For $10\,{\rm rad\,s}^{-1}\lesssim\Delta\Omega\lesssim10^{2}\,{\rm rad\,s}^{-1}$, $\Omega_{\rm gw}(\nu)$ peaks near the most sensitive part of the LIGO frequency band.  The strong scaling of $\Omega_{\rm gw}(\nu_{c})\propto\Delta\Omega^{7}$ is clear in figure \ref{OmegaGWa} and implies that the stochastic background from neutron star turbulence may be observable by ET for $\Delta\Omega$ in the above range.  From current non-detections with Initial LIGO, we obtain $\Delta\Omega\lesssim55\,{\rm rad\,s}^{-1}$.  A hypothetical Advanced LIGO non-detection with three years of data would imply $\Delta\Omega\lesssim25\,{\rm rad\,s}^{-1}$, which is high but not unphysically so.  The latter limit is competitive with the indirect spin-down limit $\Delta\Omega/\Omega\le0.04$ inferred from radio timing \cite{melatos10} for that subset of the neutron star population with $\Omega\ge7.5\times10^{2}\,{\rm rad\,s}^{-1}$.

\subsection{Shear distributions}\label{distribution}
Let us now assume that the shear is proportional to the spin-down rate, where the constant of proportionality equals the internal relaxation time-scale, $\tau_{d}$.  We consider two neutron star populations, one based on the observed spin-down rates of radio-loud pulsars, the other based on the observed spin-up rates of X-ray accreting millisecond pulsars.

\subsubsection{Radio-loud pulsars}\label{radioloud}
From the distribution of radio-loud pulsars in the ATNF catalogue \cite{manchester05}\footnote{http://www.atnf.csiro.au/people/pulsar/psrcat/}, we select objects with surface magnetic field greater than $2\times10^{10}\,{\rm G}$ to exclude millisecond pulsars, which are treated in section \ref{msp}.  We fit a log-normal distribution\footnote{The decision to fit a log-normal distribution to the data was motivated by inspection (given other uncertainties, qualitative agreement is sufficient), as well as by population synthesis models that fit, for example, log-normals to the magnetic field ($B$) distribution \cite{fauchergiguere06}, which is related to $\dot{\Omega}$ through $B^{2}\Omega^{3}\propto\dot{\Omega}$.  We quantify how good this fit is to the data by calculating the ratio of the moments of the data to that of the functional fit.  We find that the ratios corresponding to the first and second moments are both unity by construction, while the third and fourth are respectively 0.979 and 0.941.} to $\Delta\Omega=\tau_{d}\dot{\Omega}$, leaving $\tau_{d}$ as a model parameter to be constrained by future (current) gravitational wave (non)-detections.  We assume $\tau_{d}$ is the same in all objects for simplicity and is set by the viscosity for example.  Note that the $\Delta\Omega$ distribution must fall off faster than $\Delta\Omega^{7}$ otherwise $\Omega_{\rm gw}$ diverges.  Hence we cut off the log-normal distributions at a maximum shear $\Delta\Omega_{\rm max}$, related to the centrifugal break-up angular velocity, a conservative choice.  

\begin{figure}
	\begin{center}
	\includegraphics[angle=0,width=0.95\columnwidth]{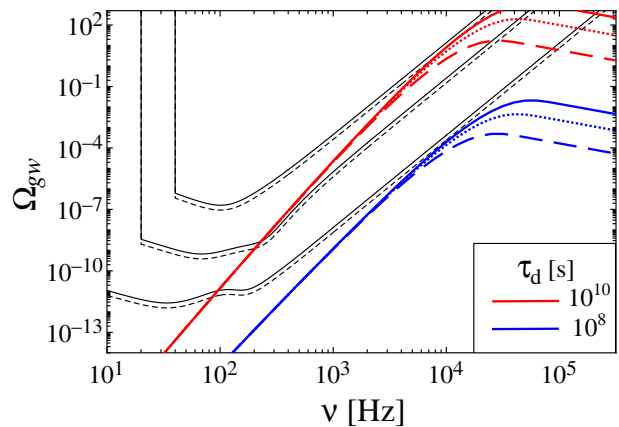}
	\end{center}
	\caption{\label{OmegaGWPulsarDistribution} Gravitational wave energy density assuming the $\Delta\Omega$ distribution for all neutron stars is identical to the $\Delta\Omega$ distribution for radio-loud pulsars, with $\Delta\Omega=\tau_{d}\dot{\Omega}$, where $\tau_{d}$ is an internal relaxation time-scale and the $\dot{\Omega}$ distribution is drawn from the ATNF catalogue \cite{manchester05}.  The solid, dotted and dashed coloured curves correspond to maximum shear $\Delta\Omega_{\rm max}/2\pi=2$, $1.5$ and $1\,{\rm kHz}$ respectively.  The black curves are the sensitivity curves for Initial LIGO, Advanced LIGO and ET assuming observation times of one year (solid curves) and three years (dashed curves).}
\end{figure}

In figure \ref{OmegaGWPulsarDistribution} we plot $\Omega_{\rm gw}$ as a function of $\nu$ for $\tau_{d}=10^{8}$ and $10^{10}\,{\rm s}$.  The solid, dotted and dashed curves correspond to $\Delta\Omega_{\rm max}/2\pi=2$, $1.5$, $1\,{\rm kHz}$ respectively.  A comparison of figures  \ref{OmegaGWa} and \ref{OmegaGWPulsarDistribution} shows that the background from the radio-loud distribution is dominated by objects with $\Delta\Omega$ near $\Delta\Omega_{\rm max}$; as $\Delta\Omega_{\rm max}$ decreases, $\Omega_{\rm gw}(\nu)$ turns over at lower frequencies and $\Omega_{\rm gw}(\nu_{c})$ decreases.  

The {\it total} gravitational wave energy density, obtained by integrating the curve in figure \ref{OmegaGWPulsarDistribution}, $\Omega_{\rm gw}^{\rm tot}=\int_{0}^{\infty}d\ln\nu\,\Omega_{\rm gw}(\nu)$, places additional constraints on $\Omega_{\rm gw}(\nu)$.  For example, $\Omega_{\rm gw}^{\rm tot}$ must be smaller than the baryon energy density inferred from cosmological observations, i.e. $\Omega_{\rm gw}^{\rm tot}<\Omega_{b}\approx0.04$ \cite{larson11}; clearly the background studied in this paper, which is emitted primarily at $z\sim1$, ultimately comes from mechanical energy in baryons created at higher $z$.  This leads to upper limits on $\tau_{d}$ that are comparable with those derived in figure \ref{OmegaGWPulsarDistribution} for the ET sensitivity curve with $\Delta\Omega_{\rm max}/2\pi\approx2\,{\rm kHz}$ and for Advanced LIGO with $\Delta\Omega_{\rm max}/2\pi\approx1\,{\rm kHz}$.

We emphasise however, that large shears, i.e. $\Delta\Omega/2\pi\gtrsim1\,{\rm kHz}$, are unlikely astrophysically, except perhaps in a small subset of young objects with super-rotating cores \cite{melatos12}.  Hence, the lesson of figure \ref{OmegaGWPulsarDistribution} is {\it not} that detections are expected at $\nu\gtrsim1\,{\rm kHz}$ from a handful of strongly sheared pulsars, but rather that (i) upper limits on $\Delta\Omega$ are best obtained from $0.2\,{\rm kHz}\lesssim\nu\lesssim1\,{\rm kHz}$, where the theoretical curves hug the detector sensitivity curves; and (ii) such upper limits on $\Delta\Omega$ are approximately independent of $\Delta\Omega_{\rm max}$. 

\begin{figure}
	\begin{center}
	\includegraphics[angle=0,width=0.95\columnwidth]{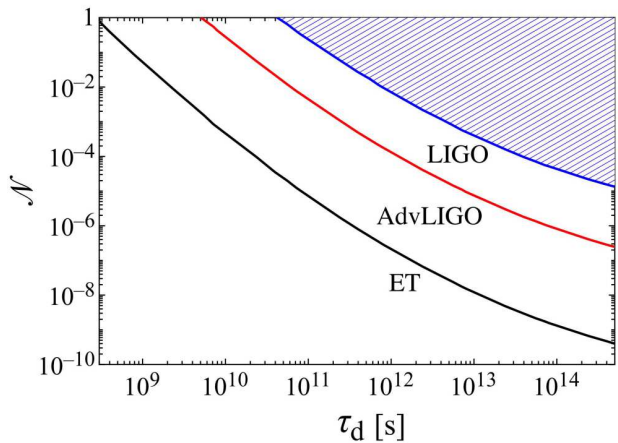}
	\end{center}
	\caption{\label{Ntau} Non-detection limits from Initial LIGO, Advanced LIGO and ET in the $(\tau_{d},\,\mathcal{N})$-plane, where $\tau_{d}$ is the internal relaxation time-scale and $\mathcal{N}$ is the fraction of neutron stars following the $\Delta\Omega$ distribution for radio-loud pulsars.  A one-year observation is assumed.  The absence of detection by LIGO so far rules out the parameter space above the blue line.}
\end{figure}

Figure \ref{OmegaGWPulsarDistribution} is drawn assuming that {\it all} neutron stars are described by the radio-loud distribution.  In reality, only a certain fraction, $\mathcal{N}$ are described by this distribution; most neutron stars lie beyond the pulsar death line, where magnetospheric electron-positron pair cascades switch off \cite[e.g.][]{arons79}.  From (\ref{Omega}), we have $\Omega_{gw}\left(\mathcal{N},\nu\right)=\mathcal{N}\Omega_{gw}\left(1,\nu\right)$, and the curves in figure \ref{OmegaGWPulsarDistribution} are upper limits; in general they are lower by the factor $\mathcal{N}$.  In figure \ref{Ntau} we plot the Initial LIGO, Advanced LIGO and ET non-detection curves as a function of $\tau_{d}$ and $\mathcal{N}$.  That is, for any ordered pair $(\tau_{d},\,\mathcal{N})$ that lies above the curve plotted in figure \ref{Ntau}, some portion of $\Omega_{\rm gw}(\nu)$ lies above the sensitivity curve of that particular detector configuration.  The shaded region above the blue line in figure \ref{Ntau} indicates the parameter space that has already been ruled out by the non-detection of a stochastic signal by Initial LIGO.  

For any given value of $\tau_{d}$, Advanced LIGO provides a limit on $\mathcal{N}$ two orders of magnitude better than Initial LIGO.  The improvement with ET is three orders of magnitude.  Alternatively, for realistic neutron star populations with $10^{-3}\lesssim\mathcal{N}\lesssim10^{-5}$, Initial LIGO non-detection implies $\tau_{d}\lesssim10^{13}\,{\rm s}$.  The limit becomes $\tau_{d}\lesssim5\times10^{11}\,{\rm s}$ and $\tau_{d}\lesssim10^{10}\,{\rm s}$ for Advanced LIGO and ET respectively.  We comment briefly on how close these limits are to challenging theoretical expectations in section \ref{conclusion}.

One can set a complementary upper limit on the combination of $\tau_{d}$ and $\mathcal{N}$ from the condition $\Omega_{\rm gw}^{\rm tot}<\Omega_{b}$ discussed above.  Again, for $\Delta\Omega_{\rm max}/2\pi\approx2\,{\rm kHz}\,(1\,{\rm kHz})$, this upper limit is similar to the ET (Advanced LIGO) curve presented in figure \ref{Ntau}.  We reiterate that such large shears are astrophysically very unlikely.  Future generations of gravitational wave detectors with higher sensitivity will place more interesting limits.

\subsubsection{Accreting millisecond pulsars}\label{msp}
\begin{figure}
	\begin{center}
	\includegraphics[angle=0,width=0.95\columnwidth]{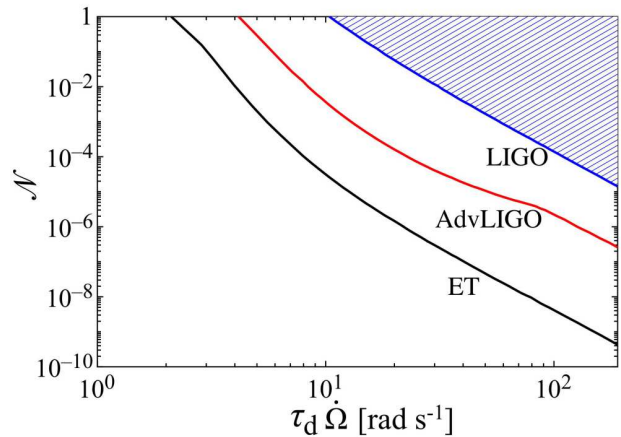}
	\end{center}
	\caption{\label{Ntaumsp} Non-detection limits from Initial LIGO, Advanced LIGO and ET for the population of accreting millisecond systems in the $(\tau_{d}\dot{\Omega},\,\mathcal{N})$-plane, where $\Delta\Omega=\tau_{d}\dot{\Omega}$ is set by the accretion spin-up torque (see text) and $\mathcal{N}$ is the fraction of neutron stars that are accreting millisecond pulsars.  A one-year observation time is assumed.  The absence of detection by LIGO so far rules out the parameter space above the blue line.}
\end{figure}

The scaling $\Omega_{\rm gw}(\nu_{c})\propto(\Delta\Omega)^{7}$ means that relatively few sources with large shears can dominate the background.  One natural place to find large shears are accreting millisecond pulsars.  The shear is maintained by the accretion spin-up torque, $N_{\rm acc}$, which is often greater than an isolated star's electromagnetic spin-down torque \cite{hartman08}.  Typically we have $N_{\rm acc}=I\dot{\Omega}\approx\dot{M}\sqrt{GM_{\star}R_{\star}}$, where $\dot{M}$ is the mass accretion rate, $I$ is the moment of inertia and the lever arm (i.e. magnetospheric radius) is approximately $R_{\star}$.  There are only a few X-ray timing observations of $\dot{\Omega}$ \cite[e.g., see][]{watts08}, and we therefore retain $\Delta\Omega=\tau_{d}\dot{\Omega}$ as the free variable that we wish to constrain.  Similar to section \ref{radioloud}, $\Omega_{\rm gw}$ is a function of two parameters: the millisecond pulsar fraction $\mathcal{N}$, and $\tau_{d}\dot{\Omega}$.  In figure \ref{Ntaumsp} we plot the non-detection curves for the Initial LIGO, Advanced LIGO and ET detectors in the $(\tau_{d}\dot{\Omega},\,\mathcal{N})$-plane.  Non-detection by Initial LIGO implies the shaded blue region is already ruled out.  

From Ref. \cite{watts08}, a typical value of $\dot{M}$ for accreting millisecond pulsars leads to $\dot{\Omega}\sim10^{-12 }\,{\rm s}^{-2}$.  Population synthesis models suggest there are $\sim10^{3}$ low- and intermediate-mass X-ray binaries out of $\sim10^{9}$ compact objects in the galaxy (e.g., \cite{kiel06,kiel08} and references therein).  Taking this as typical of the universal population, i.e. $\mathcal{N}\sim10^{-6}$, Initial LIGO non-detection implies $\tau_{d}\lesssim10^{15}\,{\rm s}$.  Advanced LIGO and ET will push this limit to $\tau_{d}\lesssim10^{14}\,{\rm s}$ and $\tau_{d}\lesssim10^{13}\,{\rm s}$ respectively.

\section{Conclusion}\label{conclusion}
In this paper we have calculated the stochastic gravitational wave background from superfluid turbulence driven by differential rotation in a cosmological populations of neutron stars, generalising the single-source calculation in Ref. \cite{melatos10}.  We found that the gravitational wave energy density per logarithmic frequency interval peaks at $\nu_{c}\sim\Delta\Omega$ and that its peak value scales as $\Omega_{\rm gw}(\nu_{c})\propto(\Delta\Omega)^{7}$.  Hence, relatively few sources with large shears dominates the background.  

We evaluated the background in three specific scenarios.  Firstly, we took all sources to have a unique $\Delta\Omega$, as when balance between the Magnus and nuclear pinning forces sets the shear in the inner crust superfluid.  It was found that the background is detectable by third-generation ground-based gravitational wave detectors such as the proposed Einstein Telescope for $\Delta\Omega\gtrsim20\,{\rm rad\,s}^{-1}$.  Secondly, we took the known distribution of radio-loud pulsars from the ATNF catalogue \cite{manchester05} to be representative of all neutron stars, the shear was then assumed to be proportional to the spin-down rate, where the constant of proportionality is the relaxation time-scale for the core and the crust to come into co-rotation in the absence of spin-down.  These models are then parametrised by the fraction of total objects that lie within this distribution, $\mathcal{N}$.  For $\mathcal{N}\sim10^{-3}$, current LIGO non-detection implies $\tau_{d}\lesssim1\times10^{13}\,{\rm s}$.  Advanced LIGO and ET non-detection limits are $\tau_{d}\lesssim5\times10^{11}\,{\rm s}$ and $1\times10^{10}\,{\rm s}$ respectively.  Finally, we calculated the background from accreting, rapidly rotating systems like low-mass X-ray binaries.  Assuming again a common relaxation time-scale, $\dot{\Omega}=2\pi\times10^{-11}\,{\rm rad\,s}^{-2}$ from X-ray timing data, and a reasonable value of $\mathcal{N}\sim10^{-6}$, non-detection by LIGO implies $\tau_{d}\lesssim1\times10^{15}\,{\rm s}$.  Advanced LIGO and ET will push the limit to $\tau_{d}\lesssim1\times10^{14}\,{\rm s}$ and $1\times10^{13}\,{\rm s}$ respectively.


The gravitational wave energy density approximately follows a piecewise power law, $\Omega_{\rm gw}=\Omega_{\alpha}\nu^{\alpha}$, with $\alpha=7$ for $\nu<\nu_{c}$ and $\alpha=-1$ for $\nu>\nu_{c}$.  LIGO and Virgo cross-correlation searches for a stochastic background have looked for power laws, albeit with $-3\le\alpha\le3$ \cite{abbott05,abbott07c,abbott09,abadie12}; see also Ref. \cite{mandic12} for a discussion of parameter estimation within these models.  It will be worth extending the range of exponents to include $\alpha=7$ in the future.  

There are a number of uncertainties in our results that require further investigation.  In Ref. \cite{melatos10}, it was shown that the root-mean-square wave strain, $h_{\rm rms}$, is insensitive to the exponent of the turbulence power spectral density.  On the other hand, the form of the eddy turnover time, $\eta(k)^{-1}$, and hence the form of the autocorrelation function, $C(\tau)$, depends more closely on the dynamics of the turbulent eddies, and these dynamics are poorly understood in superfluid turbulence \cite{salort10}.  Superfluid spherical Couette simulations modelling neutron star turbulence exhibit a Kolmogorov-like cascade \cite{peralta05,peralta06b,melatos07,peralta08}, but they are filtered spectrally to ensure numerical stability, so more work needs to be done before their output can be trusted fully.  In addition, stratification is likely to modify the turbulent spectrum (see Appendix \ref{PowerSpecApp}), and magnetic fields must eventually be incorporated too (see discussion in Ref. \cite{melatos12}).  

In light of the above uncertainties, the results in this paper are deliberately expressed as upper limits on $\Delta\Omega$ from non-detections, rather than predictions on $\Omega_{\rm gw}(\nu)$ given a known $\Delta\Omega$.  It is of course pertinent to ask how the upper limits on $\Delta\Omega$ compare with the best guesses for $\Delta\Omega$ in the literature.  One can approach this question in many ways. 
\begin{itemize}
	\item In glitches, the observed fractional angular velocity jump is $10^{-11}\le\Delta\Omega/\Omega\le10^{-4}$, although the absence of a reservoir effect, whereby the glitch size is proportional to the time elapsed since the preceding glitch, suggests this change in angular velocity is a small fraction of the underlying shear \cite{melatos08,warszawski08}. 
	\item If the angular velocity lag between the crust and the superfluid core is set by balance between the Magnus and pinning forces, the differential angular velocity can be as large as $\Delta\Omega\sim1\,{\rm rad\,s}^{-1}$ \cite{link12b}.
	\item Gravitational wave emission from hydrodynamic turbulence removes rotational kinetic energy from a neutron star, causing the star to spin down.  A fundamental upper limit follows from noting that this gravitational wave spin-down is less than the spin-down observed in radio timing experiments, giving $\Delta\Omega/\Omega\lesssim10^{-2}$ \cite{melatos10}.
	\item Buoyancy-inhibited Ekman flows create a persistent angular velocity differential between the crust and core with $\Delta\Omega/\Omega$ as high as $10^{-1}$ \cite{melatos12}.
\end{itemize}

Once a detection is made, the analysis in section \ref{upperlimits} will need to be generalized to distinguish between, and quantify the relative contribution of, different neutron star populations as well as other (e.g. cosmological) emission mechanisms.  Parameter estimation in this case is significantly more complicated and needs to be evaluated in the context of specific detection algorithms.  We do not attempt it here in view of the uncertainties outlined in the previous paragraph.  Nevertheless, by way of illustration, a cross-correlation search requires detection over a finite frequency band in order to determine the form of the power-law.  This can be complicated by the presence of multiple sub-populations of pulsars, e.g., a signal from both accreting and non-accreting systems, or the overlap of a signal from another stochastic source.  Additional complications arise by noting that the overlap reduction function is typically small at low frequencies \cite{flanagan93}, although Advanced LIGO does have some narrow-band capabilities.  The subject of parameter estimation in anticipation of future detectinos is the subject of ongoing work.

\appendix
\section{Stratified Turbulence}\label{PowerSpecApp}
Shear-driven turbulence in a fluid that is stably stratified against thermal convection is a subtle phenomenon.  Many open questions persist regarding terrestrial experiments with Navier-Stokes fluids, let alone exotic superfluids in neutron stars.  A proper treatment of stratification lies well outside the scope of this paper.  In this appendix, we flag some of the key issues briefly and point the reader to some useful references, in anticipation of further studies.  The issues are also canvassed in sections 2 and 3.3 in Ref. \cite{melatos10}.

The latest results on stratified turbulence come from large-scale (e.g. $1024\times512\times512$) direct numerical simulations in three dimensions, e.g., \cite{brethouwer07,chung12} and references therein.  The simulations are controlled by two variables: the activity parameter, $I=\epsilon/vN^{2}$, where $N$ is the Brunt-V\"ais\"al\"a frequency, and the Richardson number, ${\rm Ri}=N^{2}\left(\partial v_{\phi}/\partial r\right)^{-2}$ (or equivalently the reciprocal of the squared Froude number).  As a rule of thumb, stratified turbulence is three-dimensional for $I\gtrsim7$ \cite{iida09}, with a fully developed inertial range and dissipation at small scales.  It fossilises or ``collapses'' down to two dimensions for $I\lesssim7$, whereupon viscous shearing dominating vertical momentum transport and dissipation occurs at large scales (i.e. there is no inertial range).  Figure 18 in \citet{brethouwer07} delineates these two regimes (as well as unstratified turbulence) more precisely in the $I$--${\rm Ri}^{1/2}$ plane.  Given the scalings
\begin{align}
	I&=4\times10^{2}\left(\frac{\Delta\Omega}{1\,{\rm rad\,s}^{-1}}\right)^{3}\left(\frac{v}{1\,{\rm m}^{2}\,\rm {s}^{-1}}\right)^{-1}\left(\frac{N}{500\,{\rm rad\,s}^{-1}}\right)^{-2},\\
	{\rm Ri}&=3\times10^{5}\left(\frac{\Delta\Omega}{1\,{\rm rad\,s}^{-1}}\right)^{-2}\left(\frac{N}{500\,{\rm rad\,s}^{-1}}\right)^{2},
\end{align}
most neutron stars lie in the three-dimensional regime, near the boundary between strong and weak stratification.  Buoyancy suppresses radial motions above the Ozmidov scale \cite{chung12}
\begin{align}
	l_{O}&=\left(\epsilon/N^{3}\right)^{1/2}\notag\\
		&=0.89\left(\frac{\Delta\Omega}{1\,{\rm rad\,s}^{-1}}\right)^{3/2}\left(\frac{N}{500\,{\rm rad\,s}^{-1}}\right)^{-3/2}\,{\rm m}
\end{align}
Three other scales in the problem are the Corrsin scale, $\epsilon^{1/2}(\partial v_{\phi}/\partial r)^{-3/2}$, below which anisotropic shear production is weak, the Kolmogorov scale for viscous dissipation, $(v^{3}/\epsilon)^{1/4}$, and the radius of the star.

Even when the turbulence fossilises for $I\lesssim7$, the flow remains turbulent in spherical shells rather than in three dimensions.  This phenomenon is observed in high-resolution numerical simulations \cite{brethouwer07,chung12} and also in the Earth's atmosphere and oceans \cite{riley08}.  Large-scale, vertically sheared (i.e. streamwise elongated), horizontal motions persist under strong stratification, e.g. pancake vortices in spherical shells, and interleaved laminar and turbulent ``lasagne'' layers.  The structures are all non-axisymmetric in general.  Furthermore, they are intermittent, recurring erratically in bursts even when the system is statistically stationary (achieved in the simulations by throttling the mean shear \cite{chung12}).  Thus, a stochastic gravitational wave signal is expected under a wide range of stratified conditions, even though its character changes with $I$.  Likewise, momentum is transported vertically, whether the turbulence is two- or three-dimensional.  \citet{chung12} showed that the vertical diapycnal diffusivity scales with the activity parameter approximately as $vI^{1/3}$ divided by the Prandtl number \cite{ivey08}, with the details depending on $l_{O}/R_{\star}$.

How might the predictions in this paper change in light of the above stratification physics?  The effects enter in three places.  First, the power spectral density $P(k)\propto k^{\alpha}$ arguably changes from $\alpha=-11/3$ ($I\gtrsim7$; forward cascade) to $\alpha=-3$ ($I\lesssim7$; reverse cascade).  However, $C(0)$ changes by less than $20\%$ between these two cases \cite{melatos10}, and the scalings of $C(0)$ and $\tau_{c}$ with $\Delta\Omega$, $R_{\star}$ and $M_{\star}$ are the same for all $\alpha<-5/3$ (i.e. the signal is dominated by $k_{s}$).  Second, for $I\lesssim7$, the signal becomes intermittent (see above) albeit still statistically continuous.  Again, though, this effect is likely to wash out when observing a cosmological background in which multiple sources are superposed.  Third, stratification may invalidate our assumption of isotropy, e.g., when pancake vortices form for $I\lesssim7$.  This is a genuinely open (and difficult) question which deserves further study.  It is complicated by the fact that even unstratified high-${\rm Re}$ turbulence displays long-lived coherent anisotropic structures like hairpin vortices, e.g., in terrestrial wind-tunnel experiments \cite{ganapathisubramani03}; see also Ref. \cite{melatos10}.  Moreover, the quantum mechanical turbulence (``vortex tangle'') in a superfluid is polarised on large scales even though it is isotropic locally \cite{jou06,tsubota09} and anisotropic on intermediate scales due to patchy mutual friction \cite{peralta06b}.  Ultimately, the magnetic field must also be treated, raising other difficult issues.

\section{Wiener-Khintchine theorem}\label{app}
The derivation of equation (\ref{final}) from equation (\ref{parc}) required careful application of the Wiener-Khintchine theorem to ensure that the energy flux does not diverge but rather is proportional to the emitting time.  Our derivation follows closely that outlined by \citet{pottier10}.

Consider a stochastic process, $X(t)$, which is both real and stationary.  Realisations of this process, $x(t)$, are not square-integrable [i.e., $\int_{-\infty}^{\infty}dt\,|x(t)|^{2}$ diverges], as $x(t)$ does not vanish as $t\rightarrow\infty$.  Hence, we consider a finite time interval, $T$, and define a truncated time series
\begin{align}
	X_{T}(t)=\left\{
		\begin{array}{ccl}
			X(t) & & 0\le t\le T\\
			0 & & {\rm elsewhere}
		\end{array}\right..
\end{align}
The Fourier transform of $X_{T}$ is generally defined in terms of an integral from $-\infty$ to $\infty$ and reduces here to
\begin{align}
	X_{T}(\nu)=&\int_{0}^{T}dt\,X_{T}(t){\rm e}^{2\pi i\nu t},
\end{align}
for the truncated time series.  Moreover, the Fourier coefficients are expressed as
\begin{align}
	A_{n}=\frac{1}{T}\int_{0}^{T}X(t){\rm e}^{i\nu_{n}t}dt=\frac{1}{T}X(\nu_{n}),\label{An}
\end{align}
where the last relation assumes a fixed $T$. 

The power spectral density, $S(\nu)$, is proportional to the mean square of the Fourier transform
\begin{align}
	S(\nu)=
	\frac{1}{T}\left<|X_{T}(\nu)|^{2}\right>.\label{Sn}
\end{align}
Assuming $S(\nu)$ to be a continuous function of $\nu$, and taking the limit $T\rightarrow\infty$, one can show 
\begin{align}
	\left<X(\nu)X(\nu')^{\star}\right>=2\pi\delta\left(\nu-\nu'\right)S(\nu).
\end{align}
Evaluating this at $\nu=\nu'$ leads to an infinite energy per unit area per unit frequency: a turbulent neutron star emitting for an infinite length of time radiates an infinite amount of energy.  For a finite emitting lifetime, T, we have from (\ref{An})
\begin{align}
	\left<|A_{n}|^{2}\right>=\frac{1}{T^{2}}\int_{0}^{T}dt\,\int_{0}^{T}dt'\,\left<X(t)X(t')\right>{\rm e}^{i\nu_{n}\left(t-t'\right)}.
\end{align}
The autocorrelation function in the integrand is only a function of $\tau=t-t'$.  Integrating over the square domain $0\le t\le T$, $0\le t'\le T$, we find
\begin{align}
	\left<|A_{n}|^{2}\right>=\frac{1}{T}\int_{-T}^{T}d\tau\,\left(1-\frac{|\tau|}{T}\right)\left<X(t)X(t')\right>{\rm e}^{i\nu_{n}\tau},\label{An2}
\end{align}
and hence from (\ref{Sn}),
\begin{align}
	S(\nu)=\int_{-T}^{T}d\tau\,\left(1-\frac{|\tau|}{T}\right)\left<X(t)X(t')\right>{\rm e}^{i\nu\tau}.
\end{align}
Taking the limit as $T\rightarrow\infty$ yields the standard Wiener-Khintchine theorem that $S(\nu)$ is the inverse Fourier transform of the autocorrelation function. 

In the present application, we identify $X(t)$ with $\partial h_{jk}^{\rm TT}/\partial t$.  Hence $\left<X(t)X(t')\right>$ falls away with $\tau$ on the turbulence decorrelation timescale, $\tau_{c}$, which is of the order of milliseconds, significantly longer than the relevant emitting time ($\sim10^{9}\,{\rm yr}$).  We therefore have $T\gg|\tau|$ and hence
\begin{align}
	\left<\left|X(\nu)^{2}\right|\right>=T\int_{-\infty}^{\infty}d\tau\,\left<X(t)X(t')\right>{\rm e}^{i\nu\tau}.
\end{align}

\acknowledgments
This work was supported through an Australian Research Council Discovery Project (DP110103347).  PDL was partially supported by a University of Melbourne Early Career Researcher grant.  MFB was supported by an Australian Postgraduate Award.  We are grateful to Stuart Wyithe, Vikram Ravi, Eric Thrane and Nelson Christensen for valuable discussions about the Universe and Duncan Galloway about the neutron stars within it.  We are also grateful to the anonymous referee for suggestions that improved the manuscript, and especially for pointing out that the $\ell=2$, $m\neq2$ modes contribute significantly to the gravitational wave emission.

\bibliography{Bib}

\end{document}